\documentclass{aastex}

\newcommand{\myemail}{cabia@ugr.es}
\newcommand{\hemail}{isern@ieec.fcr.es}
\slugcomment{Submitted to ApJ}

\shorttitle{J-type carbon stars}
\shortauthors{Abia \& Isern}

\begin{document}

\title{The Chemical Composition of Carbon Stars II: \\
        The J-type stars}

\author{C. Abia}
\affil{Dpt. F\'\i sica Te\'orica y del Cosmos, Universidad de Granada,\\
    E-18071 Granada, Spain. \myemail}

\and

\author{J. Isern}
\affil{Institut d'Estudis Espacials de Catalunya - CSIC,\\
       c/ Gran Capit\'a 2-4, E-08034 Barcelona, Spain. \hemail}

\begin{abstract}
Abundances of  lithium, heavy elements and  carbon isotope ratios have
been   measured  in 12 J-type galactic    carbon  stars. The abundance
analysis shows that  in   these  stars the abundances   of   s-process
elements with respect to the metallicity  are nearly normal. Tc is not
present in most of  them, although upper limits  have been derived for
WZ  Cas and WX  Cyg, perhaps  two   SC-type rather than J-type  carbon
stars. The Rb abundances, obtained from  the resonance 7800 {\AA} Rb I
line, are surprisingly low, probably due  to strong non-LTE effects in
the  formation  of this line  in cool   carbon-rich stars. Lithium and
$^{13}$C are  found to be  enhanced  in all the  stars. These  results
together with the    nitrogen  abundances and oxygen  isotope   ratios
measured by Lambert et al. (1986) and Harris et al. (1987) are used to
discuss the origin of J-stars. The luminosity and variability class of
the  stars studied would indicate that  they are  low mass (M$\lesssim
2-3$ M$_\odot$), less evolved objects   than the normal carbon  stars,
although the presence of  some luminous  (M$_{\rm{bol}}<-5.5$) J-stars
in our  galaxy (WZ Cas   may be  an example) and  in other  galaxies,
suggests the     existence of at  least  two   types of  J-stars, with
different formation scenarios depending upon  the initial mass of  the
parent star.    Standard evolutionary  AGB  models are    difficult to
reconcile with all  the  observed chemical  characteristics.  In fact,
they    suggest the   existence  of  an   extra-mixing mechanism which
transports   material  from the   convective envelope  down  to hotter
regions where  some nuclear burning  occurs.  This mechanism would act
preferably on  the early-AGB  phase in low-mass  stars.  Mixing at the
He-core flash and the binary  system hypothesis are also discussed  as
alternatives to the above scenario.

\end{abstract}


\section{Introduction}
The carbon  content in the  envelope of  asymptotic giant branch (AGB)
stars   is  believed to   increase     along  the spectral    sequence
M$\rightarrow$S$\rightarrow$C   during   this   phase of  stellar
evolution. The   origin of this carbon enhancement    is the mixing of
He-burning   products  with    matter  from    the convective envelope
through the third dredge-up (TDU)  mechanism which can happen after
each thermal instability (pulse)  of the He-shell \citep{ibr83}. The recurrence  of
TDU episodes leads to the creation of a carbon (C) star, defined as an
AGB star with a C/O ratio higher than unity in the envelope. Among the
C-stars there exists a significant group of stars ($\sim 15\%$) named J-type
stars   \citep{bou54}   showing strong  $^{13}$C-bearing     molecule
absorptions,  which   usually implies   low   $^{12}$C/$^{13}$C ratios
($<15$) \citep{lam86,abi97,ohn99}. 

The location of J-stars in the above AGB spectral sequence is far from
clear. In  fact, some authors have  located these stars  in   a
different  evolutive sequence from that   of the ordinary carbon stars
(e.g. Chen \& Kwok 1993; Lorenz-Martins 1996), or even outside 
the AGB phase. J-stars have also been considered as the descendants of
the late-R  carbon    stars,    which have   similar     spectroscopic
characteristics \citep{llo86}. Theoretically, it is not easy to obtain an
AGB star with   the chemical  peculiarities  presented  by  J-stars.  Low
$^{12}$C/$^{13}$C ratios can be obtained in current AGB star models of
M$\geq 4$ M$_\odot$ if  hot H-burning takes  place at the bottom of
the   convective   envelope  (the    so-called   hot bottom   burning,
HBB)\citep{lat99,sac92}. However,    the  performance  of   the CN-cycle
at  the same time  destroys $^{12}$C and, in consequence, the  C/O
ratio  in   the  envelope   is  reduced and    the  star again becomes  
O-rich. Thus,  a  fine-tuning of  the parameters  of   the AGB models
(mass, mixing-length, mass-loss rate, metallicity, etc.), that determine
the chemistry of the envelope, seems to be required to obtain a J-star.  Mixing
at the  He-core flash has also been  proposed as an alternative scenario to
form J-stars. In this event an injection of  carbon-rich material from
the core  into the hydrogen-rich  shell may occur. The introduction of
core material ($^{12}$C and $^{4}$He) into a proton-rich region yields
enhanced $^{12}$C and $^{13}$C, with   perhaps a small enhancement  of
$^{14}$N \citep{deu83}. 
  
The  presence  of strong $\lambda  6708$ {\AA} Li I
lines is frequent in  J-stars.  In a Li   survey of galactic  C-stars,
\citet{bof93}  found that among  30 Li-rich stars in   a sample of 250
C-stars $\sim 50\%$ are of J-type. This figure  increases up to $\sim
70\%$ if the  Li-rich phenomenon is  considered only  among the J-type
C-stars in the survey. Although a good statistic has not yet been obtained,
Li-rich J-type  stars have also   been found in the  Magellanic Clouds
\citep{bre96}.   Interestingly,  envelope    burning   models      can
simultaneously produce Li-rich  and $^{13}$C-rich AGB  stars in models
with initial  mass M$\geq  4$  M$_\odot$. However, observations indicate  
that the majority of C-stars in  the galaxy are  low-mass objects, M$\leq 2-3$ M$_\odot$ (see,
for example, Claussen et  al. 1987). For this mass  range no  envelope burning
has been found in any AGB model. 

On  the other hand, another important
consequence of the TDU episodes in  the AGB phase is the enrichment
of the envelope with s-process  elements. These elements are  believed
to be synthesized during the period between thermal pulses, when a {\it
$^{13}$C-pocket}  (formed    in the    intershell region)    is burned
radiatively  and supplies   the  neutrons necessary    to  activate the
s-process,     via   the     $^{13}$C($\alpha$,n)$^{16}$O     reaction
\citep{str95}.  At the  next TDU, the synthesized s-nuclei  are mixed 
into the envelope.  Thus, if J-type stars owe their $^{12}$C enhancement
to the  operation  of the TDU, {\it they  should also show some  s-process
element enrichment}. There are very  few abundance analyses of J-type
stars. The  most extensive study is still  that by \citet{uts70}. This
pioneering  study based,  however, on  low  dispersion (5-14 {\AA}/mm)
photographic plates, concluded that   low-mass s-elements (Sr,  Y, Zr)
are in nearly solar proportions in  J-stars, while rare-earth elements
are overabundant by  factors of 10-100. However,   a further revision  by
\citet{uts85}  showed that in  J-type stars, abundances of s-process
elements with respect  to Fe are  nearly normal. \citet{kil75} derived
Zr/Ti ratios  in two J-stars  (Y CVn  and RX Peg)  and  \citet{dom85},
also using  low dispersion spectrograms,  defined an abundance index by
comparing the intensity of  some  lines of low-mass s-process  elements
with those of metals. Dominy drew  similar conclusions to Utsumi
with the possible exception of the  star WZ Cas. Recently, \citet{lam86}
and \citet{har87} have focussed their attention on the CNO content of
J-stars. 

The study of  correlations between different chemical species
in J-stars might cast more light upon their origin and evolution. In this work
we  perform  a detailed abundance  analysis of  twelve galactic J-type
carbon  stars using very   high  resolution and signal-to-noise  ratio
echelle spectra from the  visible to the near  infrared. To do this we
make use of up-to-date model atmospheres, and atomic and molecular data. We
focus our attention mainly  on the determination of s-process  element
abundances.    Our  results, together   with   CNO  and Li  abundances
determined in  other studies, are  contrasted with  theoretical stellar
models to find an evolutionary status  for J-type stars.

\section{Observations}
The  observations  were made  during  1997 and  1998 at  two different
observatories.  At   the Calar Alto   observatory we  used  the  2.2 m
telescope in  October-97  and July-98  during three nights in each month. For
these observations a fibre optics  Cassegrain Echelle Spectrograph was
used  (FOCES;    Pfeiffer et   al.  1998).   This   instrument uses  a
1024$\times$1024 CCD with 24 $\mu$m pixel size. The FOCES image covers
the visible  spectral region  from 0.38  to  0.96  $\mu$m in  about 80
orders with full spectral coverage.   The resolving power achieved was
$\sim  40000$ with a   two  pixel resolution   element.  At La   Palma
observatory    the 4.2  m   WHT  was   used with   the  Utrech Echelle 
Spectrograph during one night in December-97 and May-98.  The spectral
range coverage with this  instrument, using the 79.0 lines/mm grating,
is from 0.43   to 1.0  $\mu$m with  some   gaps between  orders.   The
projected size of the slit in  the 2048$\times$2048 CCD was around two
pixels (24 $\mu$m)  which gave a resolving  power of  $\sim 50000$. We
observed a  total  of 43   galactic  carbon stars   selected from  the
two-micron sky survey by \citet{cla87}. Twelve  J-type carbon stars of
this  sample  are  analyzed here  (see Table  1).  The  analysis of the
additional 31   normal (N-type) carbon  stars  will be presented  in a
future work. 

%\placetable{tab-1}
We used standard  IRAF packages and procedures  to perform bias level,
dark  current  and scattered  light  subtraction  and  to  prepare  a
normalized   flat-field  image  to   remove pixel-to-pixel
sensitivity fluctuations. A Th-Ar comparison lamp gave enough lines in
all   the  echelle orders   to     perform  an  accurate    wavelength
calibration.  We  carefully identified   non-saturated Th-Ar  emission
lines and    adjusted third-fourth order  polynomials   to  obtain a
calibration fit better than 10 m{\AA} in the residuals. The calibrated
spectra were then  divided by the  spectrum of a hot, rapidly rotating
star located in  the sky as close as  possible to  the target star  to
eliminate telluric  absorptions.  We note however,   that most  of the
lines used in the  abundance analysis are  not affected by terrestrial
features (see below). Finally,  different  images of the  same  object
were  co-added  after extraction and  calibration  to obtain the final
spectrum. The S/N ratio achieved in the final  spectra varies from the
blue to the  red   orders. At $\sim 4500$   {\AA}  the S/N  is   40-50
while at  $\sim  8000$ {\AA}  the S/N  frequently exceeded  400. In
contrast, below $\sim  4400$ {\AA} the S/N ratios  are poor because at
these  wavelengths  carbon stars  (J   and N)  are very  difficult  to
observe.  The reason for this flux depression is still a matter of controversy. In
J-stars, particularly rich  in  $^{13}$C, the absorption of  triatomic
carbon   compounds can   become  nearly  continuous  at the   shorter
wavelengths. On the other hand, \citet{joh88} have pointed out that at
the low temperatures of these stars, the resonance lines of some atoms
become so wide that they depress broad areas of the continuum. This
makes it almost impossible to observe the interesting  Tc I resonance lines
at $\sim 4250$ {\AA} in C-stars. 

\subsection{Individual stars}

The  spectral  types  shown in Table  1  are  based   on the C-system
classification by \citet{kee41}  revised  by \citet{yam77}.  Our stars
fullfil the criteria for the $^{13}$C abundance by \citet{kee93} to be
classified as J-stars:  ratio of $^{13}$C$^{12}$C $\lambda 4744$ {\AA}
to    $^{12}$C$^{12}$C  $\lambda 4737$  {\AA}   in    blue: ratios  of
$^{12}$C$^{12}$C $\lambda 6191$   {\AA} to  $^{13}$C$^{12}$C  $\lambda
6168$ {\AA}, $^{13}$C$^{12}$C $\lambda 6102$ {\AA} to $^{12}$C$^{12}$C
$\lambda  6122$ {\AA}  and  $^{13}$C$^{14}$N $\lambda  6260$ {\AA}  to
$^{12}$C$^{14}$N $\lambda 6206$  {\AA}. However, for  some of them the
spectrum is misleading, which might produce  to a problematic  spectral
classification. For instance, WZ  Cas shows weaker  CN and C$_2$ band
absorptions than the other J-stars  in the sample.  It also shows very
strong Na  D lines and its  spectrum does not  look  as crowded as the
rest of the J-stars. In fact, atomic  lines in WZ  Cas are more easily
identified. Indeed, this happens when the  C/O ratio in the atmosphere
is very close to unity, a characteristic which defines a SC-type carbon
star. Since our  C/O estimate in this  star is $\sim 1.01$, we believe
that WZ Cas has to be considered a SC-star rather than a
typical J-star. Furthermore, it   is the most luminous star  in
our sample (M$_{\rm{bol}}\sim -6.44$), which  could indicate that WZ Cas
belongs to a different population (massive) of  C-stars or that it
is in a different evolutionary status (more evolved)  than the rest of
the J-stars   in  our sample. However   \citet{lor96}   argues that  it is
difficult to  separate J-type stars and SC  stars only by spectroscopy
results.  Modeling the dusty  envelope of this  star he concludes that
WZ Cas could be a J-type carbon  star.   Less obviously,  the same
peculiarities  are  observed  in the spectrum   of WX   Cyg. In  fact,
\citet{ohn96} considered this star to be SC type although,  as far as we
know, no  other   work  in the literature    classifies  this star  as
SC-type. For this star  we do not have a  clear opinion. On  the other
hand,  FO Ser is  also classified in the literature as a late R6 type
carbon  star \citep{ste89}. We consider it a typical J-star but we
note that it presents H$_\alpha$ in absorption which is not seen in
other J stars or in  N or late R  stars. 

\subsection{Luminosities of the stars}

To estimate absolute bolometric magnitudes M$_{\rm{bol}}$ of our stars
(Table   1)  we   have  used    the empirical   relationships  between
M$_{\rm{bol}}$  and M$_{\rm  K}$-M$_{\rm  V}$ for C-stars obtained  by
\citet{alk98}.   Infrared  absolute
magnitudes  M$_{\rm K}$ are  frequently  used in connection with
distance determinations where  it is assumed  that all AGB stars
have the same  M$_{\rm  K}\sim-8.1$ \citep{fro80}.  To  obtain M$_{\rm
K}$ and M$_{\rm  V}$, the K  and V average  values (the  stars studied are
variable!) quoted in the SIMBAD database were used. K and V magnitudes
were corrected for interstellar  extinction according to  the galactic
extinction  model by    \citet{are92}.  Distances  were derived   from
parallax  measurements by HIPPARCOS.  Note  that  some parallaxes have
considerable errors (see Table 1); for instance, the large uncertainty
in   the parallax  for  UV   Cam  produces  an unlikely   value for
M$_{\rm{bol}}$ in this star. Thus,  the bolometric magnitudes in Table
1 have to be considered as average values and only indicative.  

\section{Analysis}

\subsection{Effective temperatures}
For the majority  of the stars  studied here the effective temperature
is derived by   Ohnaka  \& Tsuji  (1996,1999).  These  authors derived
effective temperatures for C-stars using the infrared flux method. The
$L'-$band (3.7 $\mu$m)  is used  as  infrared flux and  T$_{\rm{eff}}$
values  are   obtained  from  the  calibration   log$(f_{Bol}/f_{L'})$
vs. T$_{\rm{eff}}$, where $f_{Bol}/f_{L'}$ is   the ratio of  observed
bolometric flux to infrared flux. The T$_{\rm{eff}}$ values in Table 2
marked with an asterisk($^\ast$) are derived in this way. For the rest
of  the stars we   used the $(J-L')_o$ vs. T$_{\rm{eff}}$  calibration
also described by \citet{ohn96}. Infrared photometry for our stars is taken from
\citet{nog81} and  \citet{fou92}.   Effective temperatures  for J-stars
derived in this way do not  differ significantly from those derived in
N- and SC-type carbon stars. The estimated  error in T$_{\rm{eff}}$ is
$\pm 150$ K (see Ohnaka \& Tsuji 1996, for details).

\subsection{Model atmospheres}

The set of models used in  this analysis was  computed by the Uppsala
group  (see  Eriksson et al.  1984,  for  details). The models
cover the  T$_{\rm{eff}}=2500-3500$ K,  C/O$=1.0-1.35$ ranges  and all
have the same gravity  log  g$=0.0$.  The input  elemental  abundances
adopted   for  the J-star  models  were  the  solar values,  with the
exception of C, N and  O which were assumed  to be altered relative to
the Sun.  The CNO abundances in the model atmosphere  for a given star
were taken  from the literature \citep{lam86,abi97}.   For each star a
model  atmosphere was interpolated in  T$_{\rm{eff}}$ and C/O ratio in
this grid. A typical microturbulence   velocity for AGB stars  $\xi=3$
kms$^{-1}$ was  adopted or taken  from  the literature when  available
\citep{lam86}. Table 2 shows   the atmosphere parameters used  in the analysis. 

\subsection{Identification of the atomic lines}

The  greatest difficulty in  the  analysis of atomic  lines of  carbon
stars lies in the heavy blend effect of molecular  bands. The problem of
blending is even more serious in J-stars because of the strong C$_2$ and
CN  band absorptions.   For  this reason,  identification of spectral
lines  in  carbon stars  is very  difficult  and, moreover,
detailed laboratory data of these molecules are usually scarce.

\citet{uts67} carried out line  identification in two carbon  stars in
the visual region  and found that the  blend effect of molecular bands
is not so serious   in the  region   between $\lambda  4400-4500$  and
$\lambda 4750-4900$ {\AA}.  One difficulty arises, however, because at
these wavelength ranges carbon stars (particularly of J-type) are very
faint, so that very long time exposures are  needed to obtain good S/N
spectra. \citet{wal89} (7  {\AA}mm$^{-1}$) and \citet{bar94} ($\sim 3$
{\AA}mm$^{-1}$)   identified atomic    lines   in  SC-and    C-stars,
respectively,    in  a more  crowded   region ($\lambda\sim 5000-8000$
{\AA}). The accuracy in their line identifications  ranges from 0.03 to
0.08 {\AA}   depending  on the  line  intensity.  Basically,  we   have used
these three atlases for atomic line identification. 

%\placetable{tab-2}
As a  first step,  we  corrected the spectra  for the   stellar radial
velocity using the wavelengths of some easily recognized features: the
two Na  D lines, the Li I  resonance line at $\lambda 6708$ {\AA} and
the K I line at $\lambda 7698$ {\AA}, together with a few intense Fe I
and Ca  I lines. The standard  deviation around the mean Doppler shift
obtained from these  features was less than  $\pm 0.1$ {\AA}. This was
the  maximum  wavelength shift allowed  between  measured and expected
wavelengths to consider  a feature as a  good identification. Next,  we
followed the same criteria as in  \citet{abi98} (hereafter Paper I) to
consider an identification as useful for abundance analysis, namely:
first, the  line  should not present   a clear visual  blend with  any
adjacent line and second,  the {\it local}  continuum around the  line
should be  reasonably placed within   $\sim 5\%$ of  uncertainty.  Of
course,  in no way  is it possible  to establish the real continuum in
our stars.   We can only  define a local or pseudo-continuum connecting
the highest flux  points near the line of  interest. We  also tried to
use   the weakest possible   lines, log $[W(\lambda)/\lambda]\leq -4$,
although in  most of the stars  this was  not possible.  When applying
these criteria   we   immediately realized  that even   our   spectral
resolution  is not enough   to clearly  resolve   most of   the atomic
features   in the  spectra.  In fact,   the  majority  of  the  atomic
identifications in Utsumi's list  are clear blends. Therefore,
most    of  the equivalent  width  measurements   by \citet{uts70} are
overestimations, and the  corresponding  atomic  line  is not
useful   for   analysis.  This    explains    the s-process    element
overabundances derived by this author.  The atomic identification list
by \citet{wal89} and \citet{bar94}  covers  a  very crowded region   in
J-stars  and only a few  of the atomic  lines there were considered 
good candidates in our  stars. In this way we drew up a first atomic
line list.  Finally,  each  line was  checked for  possible  atomic and/or
molecular   contributions  not clearly  seen    in  the spectra   as a
blend.  When any feature  was suspected of  contributing  to the atomic
line selected, we estimated its  contribution by computing theoretical
equivalent widths  using  the model  atmosphere  parameters shown in
Table 2.  If  the estimated  contribution was  to be  higher than
$\sim 15\%$ of  the total   equivalent width  measured, the  line  was
ruled out for analysis. 

The final atomic line list is shown in Table 3. As  can be seen, very
few lines were found to be useful for analysis. Note that several 
hundred  atomic   lines were searched   in each   star using  the
above selection criteria.  For  some species only  {\it  one} line was
found. Table  3 also shows the {\it total} equivalent  widths measured. To do this
we used the SPLOT  program  of the  IRAF  package. In some  cases only
upper  limits are  derived since,  even when  the  line appears to be free  of
blends, we   believe the  uncertainty  in the  location   of the
continuum  near the line  could be  higher  than 5\%. We  estimate the
error in the equivalent width from the theoretical expression given by
\citet{cay88}  (see   also  Paper I).    The  uncertainty ranges  from
$\Delta\rm{W}(\lambda)=10$ to 35 m{\AA},  according to the line intensity
and  to the  S/N  of  the  spectrum,   with the  main  uncertainty  being
introduced by the continuum placing. 

When  possible, gf values were derived
from identification and   equivalent width measurements  in  the Solar
Atlas by \citet{mor66}, using solar abundances from \citet{and89}.  We
used  the   solar model atmosphere    by  \citet{hol74} with parameters
T$_{\rm{eff}}=5780$ K, log  g$=4.44$ and microturbulence variable with
optical depth.   Note that  the  possibility  of using   the Sun  as a
standard to derive astrophysical gf-values   is greatly limited by  the
weakness  of the   lines used  here  in the   solar spectrum. Specific
references   for individual lines are  given  in Table 3, otherwise we
used the gf-values given in  the VALD database \citep{pis95}. Finally, we considered
the hyperfine    structure using   such  information    as was  available:
\citet{mac98} for Ba: for Tc and Rb  lines see the discussion in Paper
I  and  references therein.  Broadening by radiation  damping was
calculated      as in \citet{edv93},     when not  given explicitly by
VALD. Finally,  the classical  van der Waals   damping constant of the
atomic lines was modified, also following Edvardsson et al.

\section{Abundance results}
\subsection{Lithium}
All our   stars   show   a  strong  $\lambda  6708$   {\AA}   Li I
absorption (see Figure 1).  In  WZ  Cas and  WX  Cyg  the equivalent  width  of this
spectral feature    is certainly larger  than  1  {\AA}. \citet{abi99}
discuss the formation of Li I lines  in C-rich atmospheres considering
also N-LTE effects. They conclude that among the Li lines available for
analysis in AGB stars ($\lambda 4603$,  $\lambda 6104$, $\lambda 6708$
and $\lambda 8126$  {\AA}),  the subordinate  line  at $\lambda  8126$
{\AA}  is probably the most reliable,  on the basis  of  high S/N
spectra and  spectral synthesis. This   line forms deep  enough in the
atmosphere  where the uncertainties in  the model atmosphere structure
of AGB stars are  smaller. N-LTE effects for  this line are also weak ($\sim
0.2$  dex)  and the continuous  opacity  coefficient seems  to be well
reproduced    by    model      atmospheres   in    this     wavelength
range. Unfortunately,  in  our stars,  except WZ Cas   and WX Cyg, the
$\lambda 8126$  {\AA} Li line is weak  and very crowded with strong CN
absorptions in such a way that synthetic fits  to this line differing
by $\sim 0.2-0.3$  dex in the  Li abundance do not significantly alter
the quality of the fit.  Thus,  we decided  to  use the $\lambda  6708$
{\AA}  line,  which is  much  more sensitive to  abundance variations, 
except for the  stars WX Cyg and  WZ Cas. For these  two stars we used
instead the  $\lambda 8126$ {\AA}  Li I line.  In any case, all the Li
abundances were  derived by spectrum  synthesis  and corrected by N-LTE
effects  according  to  \citet{abi99}  (see  this   paper for
details). However, one has to be very careful when interpreting the Li abundances
derived from the resonance Li I line. The presence of a circumstellar component 
in the $\lambda 6708$ Li absorption would lead us to overestimate the Li abundance 
when derived from this line. In fact, VX And, BM Gem and V614 Mon
show weak blueshifted Na D line absorptions, probably indicating the presence of
a circumstellar envelope in these stars. This circumstellar absorption is,
however, not observed in the strong $\lambda 7698$ {\AA} K I resonance line which
should form at about the same depth in the atmosphere than the Li resonance
line (see Barnbaum 1992). The lack of K I circumstellar absorption probably rules
out significant contamination of the photospheric feature.

Final Li abundances are shown  in Table 4  in the scale log
$\epsilon$(Li)$=12 +$ log(Li/H), where Li/H is the  abundance of Li by
number. From Table 4  it is clear that all  the stars have unusual  Li
abundances  (log $\epsilon$(Li)$\gtrsim 1$),   larger than the typical
value found in normal C-stars (log $\epsilon$(Li)$\sim 0.0$), but
significantly smaller than those found in the so-called super Li-rich. WX 
Cyg and WZ  Cas are certainly super Li-rich stars, although these stars may not 
be J-type stars (see above). The formal uncertainty in Li abundances of 
Table 4 ranges from 0.3-0.4 dex. Figure 2 shows the correlation of Li
abundances versus $^{12}$C/$^{13}$C ratios found in J- (this work) and
N-type carbon stars \citep{abi97}. J-type stars are all Li-rich and note
that there are also some Li-rich N-type stars. The formal error in the
$^{12}$C/$^{13}$C ratios in Figure 2 is $\pm 6$ (see Abia \& Isern 1997). 

\placefigure{fig1}
\placefigure{fig2}

\subsection{Technecium}
The presence of Tc in the atmosphere of  AGB stars (in fact $^{99}$Tc)
is commonly interpreted as evidence of  the operation of the s-process
within stars. This study is the first  detailed search for the presence
of   this element  in  J-type stars.  As mentioned   above,  the three
resonance   Tc I lines near  $\lambda  4250$  {\AA} are unaccessible in
C-stars  because  of the  strong flux  depression  in  these stars  below
$\lambda\sim 4400$ {\AA}.  Therefore we have used, as  in Paper I, the
intercombination and weaker  Tc line at $\lambda  5924.47$  {\AA}.  We
followed the same procedure in the analysis as in  Paper I. The reader
is  referred   to  this  paper   for  a  detailed  discussion    of the
identification of the  Tc blend  and the  choice of the
gf-value and the hyperfine  structure  of this  line.  As in Paper I,   the
$\lambda 5924$ Tc  blend is not well  reproduced by synthetic spectra,
in  particular the red-wing  of the line.  Thus, although in some stars
the Tc blend was apparently well reproduced, we  prefer to be cautious
and record the Tc abundance as  equal-to-or-less-than. Figure 3 shows
a clear example of this in the star WZ Cas. Tc may  be present in this
star but the poor fit to the red part of the line prevents us from asserting
a definitive     detection.   We  place an     upper   limit   of  log
$\epsilon$(Tc)$\lesssim 1$. For WX  Cyg, the other possible detection,
we set log $\epsilon$(Tc)$\lesssim  0.7$. Note that these two possible
detections are of the same level as the  Tc upper limits set for the
sample of SC stars analyzed in Paper I. In  the remaining stars, the best
fit  to the  Tc blend is   compatible  with no-Tc,  i.e.  a synthetic
spectrum with no-Tc  does not differ from another  one computed  with a
small Tc abundance. For these stars we quote a {\it no} entry in Table
4, meaning that Tc, very probably, is not present. An example of this is
shown in Figure 3 for the star UV Cam. In  fact, for this star, Y CVn
and  RX Peg we were able to obtain spectra  in the $\lambda  4250$ {\AA} region
with a high enough  S/N ratio to check  for the presence  of the Tc resonance
lines. We have not quantitatively analyzed these spectra because we lack
the appropriate atomic and  molecular lines in  this spectral region,
but a careful search for the resonance lines confirmed the absence of
technetium in these three  stars. We  agree with the  previous
finding  by \citet{lit87} in  Y CVn. \citet{bar91} report the possible
detection of Tc in two J-stars, EU And and BM Gem (star studied here). Their
argument is based on the presence of a strong absorption at $\lambda 6085$
{\AA} which is partially due to the $\lambda 6085.22$ {\AA} Tc I line. From
these authors, the fact that the $\lambda 6085$ absorption appears which similar
intensity only in those carbon stars where Tc has been detected unambiguously
using the blue lines, supports the identification of the feature at
$\lambda 6085.22$ {\AA} as Tc. Our quantitative analysis of the $\lambda 5924$
Tc feature in BM Gem is, however, compatible with a non-detection. We note that
the $\lambda 5924$ Tc feature is a factor $\sim 2$ more intense than the $\lambda 6085$
one \citep{gar81}, therefore the presence of Tc in BM Gem in any measurable amount
should have appeared in our analysis. Furthermore, the $\lambda 6085$ Tc feature
is strongly blended with a Ti I line of moderate intensity ($\chi=1.05$ eV, log gf$=-1.35$)
and some CN and C$_2$ absorptions, which necessarily requires a spectral synthesis
analysis to confirm the detection. Leaving  apart the possible presence of Tc
in BM Gem for a further and accurate analysis and the upper limits
set for WZ Cas and WX Cyg, possible SC-type stars, we can conclude
that most of J-stars do not show Tc.

\placefigure{fig3}

\subsection{Rubidium}
The Rb  abundance is a monitor  of  the neutron  density at  which the
s-process operates   in  AGB stars.  Therefore   the  derivation of Rb
abundances in  these  stars is extremely  important,  specifically the
abundance ratios  between  Rb and its  neighbors in   the periodic
table  (Zr, Sr). We have used  the resonance line at $\lambda 7800.23$
{\AA}  to derive Rb abundances. The  other  accessible line at $\lambda
7947$ {\AA} is  much  weaker and very  crowded with  CN lines in  cool
stars.  Nevertheless, the resonance line  is  also blended, and so Rb
abundances have to be derived from  spectral synthesis.  We have used
the same atomic and  molecular line list in  the Rb spectral region as
in  Paper I  with  the addition of   some  C$_2$ lines (including  the
$^{13}$C isotope) computed  by P. de Laverny  (private communication).
We refer again to Paper I for a discussion  of the identification of the
atomic and molecular  lines contributing to  the  $\lambda 7800$ {\AA}
blend. The  Rb line  is  represented  by the  hyperfine    structure
components  of both isotopes $^{85}$Rb and  $^{87}$Rb in a terrestrial
ratio    ($^{85}$Rb/$^{87}$Rb=2.59)   with gf-values     taken    from
\citet{wie80}.  Only in three stars (WX Cyg, WZ  Cas and V353 Cas) does the
Rb line  appear clearly as a prominent  absorption in the background
of CN lines.  In the remaining stars  the Rb line  is not distinguished
from the background of lines (see Figure 4). Table 4 shows the
Rb  abundance derived  in our  stars   relative  to their  mean
metallicity  [M/H]\footnote{We    adopt  here  the   usual    notation
[X]$\equiv$log(X)$_\star$-log(X)$_\odot$  for  any  abundance quantity
X.}. We adopt the solar photospheric Rb abundance by
\citet{and89}: log $\epsilon$(Rb)=2.60$\pm 0.15$. If the meteoritic Rb
abundance  is preferred, the  [Rb/M]  values in  Table   4 have to  be
increased by 0.2 dex. 

From  Table 4  it   is apparent  that the   [Rb/M] ratios derived  are
remarkably low. For  some stars the best  fit is  compatible with {\it
no} Rb. Nevertheless, we believe that our  Rb abundances could be, and
in  some   cases  are,   lower  limits.  The   are several reasons for  this:
first, some metallic lines (Fe,Ni) near  the Rb line are best
fitted by   theoretical  spectra assuming  abundance  values which are
systematically lower   by $\sim 0.2-0.3$  dex  than  that of  the mean
metallicity of the  star derived from other  metallic lines (see Table
3). Thus, the [Rb/M] ratios should be  increased by this factor if
[M/H] is derived from the atomic lines near Rb. On the other hand, our
synthetic spectra typically give a very strong $\lambda 7800$ {\AA} Rb
I absorption even for very low Rb abundances. We  found the same figure
when deriving abundances  from other resonance  or very low excitation
energy lines of elements with a similar ionization potential to Rb (4.18
eV). Consider, for instance, the resonance  $\lambda 6708$ {\AA}  Li I (5.39 eV)
and $\lambda 7698$ {\AA} K I (4.34 eV) lines, two alkali elements with a
similar atomic structure. In fact, synthetic fits to the K I line give
low potassium abundances (undersolar).   No nuclear mechanism  able  to
destroy potassium in stars is known. This effect with K however, was not found
by  \citet{ple93}   in M  giants  of  the Magellanic  Clouds.  Non-LTE
effects (overionization) could be  the reason for this as in the
case   of  the  strong resonance     Li  I  line  in   some   C-stars
\citep{abi99}. N-LTE corrections on this line  might extend to until +0.6
dex in  the sense of N-LTE minus  LTE abundances. Similar unexpected  low [Rb/M]  
ratios were also
derived by \citet{ple93}   and \citet{lam95}.  However, these  authors
conclude that NLTE  effects on Rb  are probably weak, from  analysis of
the Rb  line in Beltegeuse  and Aldebaran, two  galactic M supergiants
whose atmospheres are presumed to be similar to those of the O-rich AGB
stars they studied.  A quantitative N-LTE  study of the formation of
the Rb line in cool  C-rich atmospheres is  needed before this
question can be answered. Typical maximum uncertainties  in the  atmosphere parameters
($\Delta  \rm{T_{eff}=\pm   200}$ K; $\Delta\xi=\pm  1~\rm{kms}^{-1}$;
$\Delta  \rm{CNO/H}=\pm 0.3$ dex;  $\Delta \rm{C/O=\pm 0.05}$ dex) and
$\sim 5\%$ in the  continuum added quadratically, represent  a maximum
uncertainty  of  $\sim  0.4$  dex in  the  absolute  abundance  of  Rb
derived. The  formal uncertainty concerning  the [Rb/M] value is probably less
than this, because  some of these  sources  of uncertainty affect  the
[M/H] values in  a similar way. Thus,  when deriving the [Rb/M] ratio,
many errors would be canceled out. However, we avoid to estimate neutron
densities in the s-process site from the derived Rb/Zr or Rb/Sr ratios
due to the uncertain Rb abundances as mentioned above.

\placefigure{fig4}

\subsection{Metallicity and Heavy Elements}
The abundances of metals were derived from the usual method of equivalent width 
measurements and curves of growth calculated in LTE. Ca, V, Fe and Ti abundances 
were used  as a measure of the metallicity
of  the stars. The  [M/H]   value  shown  in  Table  4  is the   mean
metallicity obtained from these elements. The upper limits in Table 3 were
not considered when deriving [M/H].    In the  star Y  CVn  we  were not
able to identify  any metallic line useful for  abundance analysis. For this
purpose, we adopted the metallicity obtained by  \citet{lam86} from several Fe and
Ca lines. Note that the number of metallic lines analyzed per star is
rather low: minimum, one  and maximum, eight for WZ  Cas. This star is
the only one for which a reasonable statistic  with Fe lines (five) can
be performed. We found a mean dispersion  of $\pm 0.1$  dex around the mean
iron abundance derived, which is compatible  with the error introduced
by the uncertainty in the  equivalent width measurements. On the other
hand,  the  elements having isotopes  formed  by neutron captures have
very  few useful lines in  the  visible spectra of  J-stars. Note  the
significant number of empty entries or upper limits in Table 3. WZ Cas
is again the  sole star where it is  possible  to detect a significant number  of
heavy element  lines.  A resolving power  of $\sim  10^5$ is needed to
perform an   accurate analysis  of   these   stars. This means   that
abundance  analyses  in  C-stars based   on intermediate-low resolution
spectra and/or on the  visual intensity of spectral  lines can lead to
important  errors. For  example, the   $\lambda  4607.34$ {\AA}  Sr I,
$\lambda 4554.04$ {\AA}  Ba  II  and $\lambda   6709.49$ {\AA}  La   I
features, used by \citet{dom85} to  define an abundance index of these
s-process   elements,   appeared in   our  spectra   as  very  crowded
blends. At  these wavelengths  many  CN and C$_2$ lines  
contribute significantly in  C-stars. Therefore, a high intensity of
such lines does  not necessarily mean  an enhancement of Sr, Ba  or
La.  This kind  of analysis  is  only  useful in relative   abundance
studies between stars, not to derive absolute abundances. 

Figure 5 compares the strength of the Ba II line (4934.07 {\AA}) found
in three J-type stars. To establish a wider comparison, the spectrum of
a normal  (N)    C-star (Z  Psc),   presumed  to be  rich   in Ba
\citep{uts85}, is included.  All the spectra in Figure 5 were obtained
in the same way with echelle spectrographs. The  Ba II line is clearly
strong in Z  Psc, reflecting its probable  Ba overabundance,  but among
the J-stars this line  is not so  intense. Since in general few  lines
are available,  it  is not practical to  examine  the distribution  of
abundance against atomic number  or  even the mean  difference between
low-mass (Sr-Y-Zr) and high-mass  (Ba-La-Ce-Nd-Sm) s-elements  star by
star.  Instead, we derive the mean heavy-element enhancement [$<h>$/M]
shown in Table 4. To derive this we did not consider upper limits or
the   uncertain Rb  abundances.  

As in  Paper   I, an estimate was made of  the
theoretical  errors concerning the derived   metal   abundances  due to the
uncertainties in the atmosphere parameters  of the stars. The
formal error due  to  errors in T$_{\rm{eff}}$,  microturbulence,  CNO
abundances, the    C/O  ratio  in the   atmosphere,   equivalent width
measurements and the  location of  the continuum, added  quadratically,
is $\pm 0.3-0.6$ dex, depending mainly on the intensity and
excitation energy of  the line as well as  on the ionization  state of
the line considered. The microturbulence parameter is also an important
source of uncertainty since most of the abundances are derived from
strong lines, in the flat part of the curve of growth. For example,
a variation of $\Delta\xi=\pm 1$ kms$^{-1}$ produces a change of
$\pm 0.4$ dex in the barium abundance derived from the strong Ba II
line at $\lambda 4934$ {\AA}. Unfortunately, given the large uncertainties in the
equivalent width measurements and the few number of lines identified (\S 3.3), it was
impossible to estimate $\xi$ using the requirement that individual abundances derived 
from lines of different intensity have to be nearly equal.

Taking the error bar above into account our results in
Table 4 show that   J-type  C-stars  are  of near   solar  metallicity
$\overline{\rm{[M/H]}}=0.12\pm 0.16$ and {\it do not show} the sizable
heavy  element    enhancements     typical  of   S     or   SC   stars
\citep{smi90,abi98}. The  mean heavy   element enhancement  among  the
J-stars in the sample is [$<h>$/M]$=0.13\pm 0.12$, which is compatible
with {\it  non-enrichment}. Considering  individual stars (see Table 4), in  some of
them there is a hint of a heavy-element enrichment, but given
the small number of lines analyzed and the  large errors, this has to be
considered with caution. 

\placefigure{fig5}

Comparison with the results obtained by \citet{uts85} for the stars in
common (UV Cam, WZ  Cas, Y CVn and   RY Dra) is difficult  because of the
different  methods  used in  the analysis. Furthermore,
Utsumi uses only Ti  as the metallicity   indicator and refers  all the
abundances to this element. Instead, we have used Fe, Ca,  V and Ti to
obtain   the metallicity [M/H]. However,   the abundance ratios found
here agree with those of Utsumi in the stars in common between
the error bars (Utsumi estimated an accuracy of about 0.4 dex). 

\section{Evolutionary Considerations}

J-type carbon stars  are not rare, amounting to  $\sim 15\%$ of C-rich
giants, and therefore they should represent  a stage of evolution that is
available to a  significant fraction of stars,  and are not the result
of anomalous initial  conditions or statistically unlikely events. The
chemical abundances found  in   the present  and  other studies  offer
constraints to certain scenarios that have been offered to account for
the existence of these C-stars.  In the following paragraphs, we use these abundance  results
to discuss the evolutionary status suggested and to propose new ones. 

First, let us recall the chemical properties  of J-stars: a) they are
certainly carbon stars (C$>$O) and show very low carbon isotope ratios
($<15$). In many  of them the $^{12}$C/$^{13}$C  ratio is equal to the
CNO cycle equilibrium  value.  b) An important fraction  ($\sim 75\%$)
have enhanced  Li (log $\epsilon$(Li)$>1$)   although the majority are
Li-rich rather than    super   Li-rich objects.  c) They    are  solar
metallicity stars.   d)  They do  not show  Tc   or s-process element
enhancements   in their  atmospheres.  Obviously,  exceptions  to these
figures can be  found but we are only discussing these stars on the  basis of their
most common   properties. For instance,   the carbon  star  UV Aur (a symbiotic star) is
classified as  J-type, although it shows  Tc \citep{lit87} and   does not present Li
\citep{bof93}. Unfortunately, we could not   include this star in  the
present study.  

Figure  6  shows the position of  our  J-stars in an observational H-R
diagram,  including some galactic  R-type  and N-type carbon stars with
absolute   magnitudes also derived     from the HIPPARCOS  parallaxes (see
Alsknis et al. 1998). From  this figure, one might  consider
J-stars as  transition objects between   R-stars and N-stars. This  is
reinforced considering the fact that most J-stars are irregular or
semi-irregular variables (very few  Miras  are found among them)  with
not  very  large pulsation periods,  which  is a characteristic of the
less evolved carbon stars.   Furthermore, on average the   envelopes of
J-stars are thinner than those of ordinary carbon stars \citep{lor96},
which could also  suggest that  this class  of objects is  in the very
early    stages of  carbon star    evolution.    

\placefigure{fig6}

Current AGB   models
\citep{sac92, lat99,  blo98} can obtain  C-rich  (C$>$O) envelopes and
low carbon  isotope ratios   in stars  with initial mass   M$\gtrsim 4$
M$_\odot$   through the successive   He-shell  flashes and TDU  episodes
coupled  with  the  operation  of HBB  at the  base  of the convective
envelope.  These stars can also be,  for a long  period of time, Li-rich
AGB    stars  with   peak   Li     abundances  in  the     range   log
$\epsilon$(Li)$\simeq 3-4$. However, the operation of HBB leads to the
transformation of $^{12}$C into $^{14}$N; thus nitrogen is expected to
be enhanced in  the envelope of  these stars. The nitrogen  abundances
derived in some J-stars (Lambert et al. 1986) show a normal N/O ratio,
much   lower  than  that expected    on the basis   of  the  CNO cycle
operation in HBB.       The $^{17}$O/$^{16}$O  and   $^{18}$O/$^{16}$O ratios
measured in J-stars (although with an important error bar) also argue
against  a pure CNO  cycle interpretation  of the J-stars chemical
{\it anomalies} \citep{har87}.   Models by \citet{lat98}  can only obtain a
C-rich, Li-rich, $^{12}$C/$^{13}$C low and N/O$<1$ AGB  star in a
very narrow range  of stellar masses (M$\sim  5$  M$_\odot$), with a
specific   metallicity (Z$\sim$ Z$_\odot$/3)  and for a very short  
period  of time ($\lesssim 10^4$  yr).   In this context, the 
number of J-stars expected would be very  low, which is  in contrast with  the
significant number observed. On  the other hand, these objects would
be fairly luminous  (M$_{\rm{bol}}<-6$), and   should present  some
s-process  element enhancement  (at least  of  low-mass Sr-Zr-Y,  see
Vaglio et al. 1998). None   of this is   observed (see Tables 1  and
4). Thus, standard AGB  models for masses  M$\gtrsim 4$ M$_\odot$  are
very  difficult   to reconcile with      the observed properties   of
J-stars.  Note in   addition   that there is observational    evidence
indicating  a low-mass (M$\lesssim  2-3$  M$_\odot$) progenitor star for
most of the J-stars studied here (e.g. Claussen et al. 1987). Most
of our stars are not very luminous objects (M$_{\rm{bol}}>-5$). Their
luminosity is of the order of the  predicted value for low-mass stars
during or near the  AGB  phase. 

Although the formation of low-mass C-stars  has recently been found to
be possible \citep{str97}, HBB is not found in low-mass AGB models
because of    the  low  temperatures reached   at  the  base   of the   convective
envelope. One has   to advocate,  therefore,  the  existence  of a
non-standard mixing mechanism which transports material from the bottom
of the  convective envelope into  deeper and hotter regions (basically
the H burning shell) where  {\it cool  processing} might occur. This
hypothetical mixing mechanism, perhaps induced by meridional currents,
was proposed by \citet{was95} and  has been shown to reproduce the CNO
isotope  anomalies found in  some  low-mass red  giants (Boothroyd  \&
Sackmann  1999). Under certain conditions,  it  can also create $^7$Li
via the Cameron and Fowler mechanism \citep{cam71}, thus accounting for the
recent discovery  of   surprisingly high  lithium  abundances in  some
low-mass red giants    \citep{bro89,  wal82, del97}.   Boothroyd    \&
Sackmann (1999) suggest   that  this  extra-mixing and  cool    bottom
processing could also occur in low-mass  AGB stars and account for the
$^{18}$O  depletion  and low $^{12}$C/$^{13}$C    ratios found in  the
J-stars  analyzed   by Harris et  al. (1987).    The operation of this
mechanism on the early-AGB or just after the onset of the helium shell
flashes  is preferred  since  cool bottom  processing  appears to
become weaker as the star  ascends the AGB phase. This point
might  be compatible with the suggestion  (see  Figure 6) that J-stars
are not very  evolved AGB  stars but are just on  the verge of   becoming
normal N-type carbon stars.  In that case, little or no s-process
element enhancement would be expected, as a  significant number of TDU
episodes are needed. This might also be compatible with the abundance results presented
here. A quantitative analysis of the operation of this extra-mixing and
cool bottom processing and the subsequent envelope  chemist on the AGB
is currently in progress (Sackmann \& Boothroyd 1999, in preparation). 

With   the  above AGB scenarios encountering difficulties in
explaining the existence of J-stars, we examine  a scenario outside the
AGB phase: the mixing at the  He core flash.   This mechanism has 
already been proposed to explain   the evolutionary status  of the R-stars
(Dominy 1985). Note that as far  as the chemical composition is concerned,
R-stars and  J-stars   only  differ in  the   presence of   Li  and slightly
lower $^{12}$C/$^{13}$C ratios in the latter. Thus, there is the suspicion that  
R-stars may be the ancestors of J-stars (see Figure 6) after an additional 
mixing event (second dredge-up?). Very few studies   of  the core He-flash have been 
published.  In an  off-centre core  flash hydrodymanic calculation,   Deupree \& Cole
(1981, 1983)  show that a bubble of  low  density and high temperature
(T$\sim  4\times  10^8$ K) can form  and  mix the He-core  and H-shell
matter. If in some red giant stars  the core flash does introduce C-rich
material into the H-burning  shell, there is reason to believe that
the results  of He-burning and  CN-processing may reach the convective
envelope and ultimately the surface. The main question is whether the
He-flash is able to produce and  mix enough $^{12}$C into the envelope
to transform the star into  a C-star. Investigations by  \citet{men76}
of non-central flashes  (in  fact, near  the  core boundary)  within  a
rapidly rotating  core show that a  series of  flashes could occur and
build up the C  abundance in the envelope through successive flash and mixing
events.    Recently, \citet{deu96} have re-examined the He-core flash performing
hellium flash calculations of different intensity. The violence of the
flash is mainly
governed by the degree of degeneracy where the explosion occurs. The authors
estimate the surface abundance anomalies produced by the different He-flashes. 
They show that the primary material mixed into and
above the hydrogen shell in all cases is $^{12}$C. The other major
products are the result from hot $\alpha$ captures that occur during the flash
($^{20}$Ne, $^{24}$Mg, $^{28}$Si and $^{32}$S) and, if the hydrogen
shell is penetrated at reasonably high temperature, some $^{14}$N. Observable
enhancements of $^{12}$C in the envelope are favoured in very metal poor,
low-mass envelope red giants. For moderate peak temperatures
of the flash ($\sim 9\times 10^8$ K) important $^{12}$C enhancements
(by a factor of 70) can also be obtained in solar metallicity stars, in
such a way that the star might become a carbon star.
However,  in order  to   account for  the observed chemical
peculiarities   of     J-stars,  fine tuning     of    the models is
required. First,  on the way, $^{12}$C must  be  exposed to  protons  and  much  
of it be converted  to $^{13}$C to   get a low  $^{12}$C/$^{13}$C ratio   
in the envelope. In addition, to prevent the  release of neutrons via the $^{13}$C($\alpha$,n)
reaction  and so the creation of  s-process nuclei, the temperature in
the  mixing zone  must  not exceed $\sim 10^8$  K (interestingly, Deupree \& Wallace
claim that their flash computations do not produce s-process elements). Finally, Li
production would   require temperatures   not exceeding  $\sim 5\times
10^7$   K in the processing zone.   At least the last temperature requirement
appear  rather difficult to attain   (Lattanzio, private
communication). Perhaps, Li can be produced after the He-flash by an additional
mixing event between the convective envelope and the H-shell. Obviously, the He-flash  
scenario merits further hydrodynamic (3-D) studies.   Note, the He-flash  occurs in
low-mass and low-luminosity objects such the stars studied here. 

Finally, we consider the mass-transfer scenario in a binary
system. The existence   of S stars with   no  Tc, as   predicted by the
mass-transfer paradigm, is now  well established \citep{jor93}. Whether
this scenario   can also be applied to    C-stars is not  yet  firmly
demonstrated, although \citet{bar93} found 16 Tc-poor stars in a sample
of 78 C-N stars with Ba excess. However, it is difficult to
explain the  absence  of  s-process  element enhancement and  the  C/O
ratios in  our stars within this  scenario. In principle, the accreted
material  must  be extremely   carbon-rich; the donor star should be a
normal C-star with probably enhanced s-nuclei in the envelope. \citet{dom85}  
estimated  a C/O$\gtrsim  5$ in the  material accreted by a  {\it typical} 
$\sim 1$ M$_\odot$ red  giant when applying  this scenario to  
explain the C/O$>1$ ratios observed in R-stars. The same figure can be 
applied in the case of J-stars. This  extreme C/O ratio  is  not observed in  
any C-star. Furthermore, even  assuming  that   the  material transferred  were
Li-rich (some N-type carbon stars are Li-rich), it is unlikely that Li
could survive during the mass-transfer and  posterior mixing. In fact,
extrinsic (binary, no Tc) S stars do  not usually show the Li enhancements
found here \citep{ba92}.  

Nevertheless, a significant number of J-stars ($5\%-10\%$, LLoyd-Evans 1991) 
show a very uniform 9.85 $\mu$m emission which is believed to be due to the
presence of a silicate dust shell \citep{lit86}. Also the detection of
H$_2$O masers in five J-type stars has been reported \citep{eng94}. 
This is rather strange, because  silicate   emission and H$_2$O masers are   
usually  associated    with O-rich environments, while J-stars are C-rich  objects.  
It is difficult to interpret these observations. The circumstellar shell is assumed to be
produced by mass loss from the central  carbon star and should reflect
the  carbon-rich material of its    photosphere. The transit  time of
material through the circumstellar shell is  only of the order of years,
which  is   shorter  than the    transit  time ($\sim 10^5$  yr)  from
M$\rightarrow$S$\rightarrow$C.  Thus, it    is unlikely  that  we  are
observing  very  recently  produced   circumstellar  material from   a
progenitor M star. Our stars were classified as J-type C-stars over 50
years  ago.   It has  been  suggested   \citep{lit86, llo90, llo91}  that the
material  expelled from the now carbon   star, starting while it still
had  an  oxygen-rich envelope,  has  accumulated  in a   disc (or common envelope)
around an unseen hypothetical companion. \citet{lam90} discuss in detail the
different possibilities for the nature and mass of the companion star, although the 
most probable situation is that the secondary is a low mass star on the main 
sequence. In fact, \citet{bar91} found radial velocity variations by 6 km s$^{-1}$ 
in BM Gem and by 5 km s$^{-1}$ in EU And over a 6 month interval. The small uncertainty 
in their radial velocity measurements ($\pm 1.5$ km s$^{-1}$) and the fact that the
velocity has change of direction over this period of time, point out to a
binary nature for these two J-type carbon stars. Although the binary hypothesis
can probably explain the silicate emission in some J-stars, unfortunately there are 
not other radial velocity variation studies nor a search for ultraviolet excesses  
(in the hypothesis that the companion is now a white dwarf) to test the binary  
scenario for all the observed J-stars. Note that  \citet{macl97} found no evidence
of binary motion in a sample  of 22 R-type C-stars (possible ancestors
of J-stars). He however,  concluded that since  it is very common  to
find  binary  systems  ($\sim 20\%$)  among  normal  late-type giants,
it is likely that the  R-stars were once all  binaries, but with separations so
close that would not allow them to evolve completely  up the giant and
asymptotic   giant branches without coalescing.   The star with the smallest
mass  in the  system was  disrupted and engulfed   by the now visible
J-star. Perhaps  a strong non-standard  mixing event between 
the core and the envelope material   at  the He-flash was induced  by tidal forces in 
the actual J-star. Whether  this   stellar merging  or non-standard He-flash triggered
by binarity are able   to  induce  the chemical properties observed in J-stars 
is an open question.

\section{Concluding Remarks}
Our most important conclusion is that heavy element abundances in J-type
carbon stars are nearly solar with respect to their metallicity. We did not
found Tc in these stars, although we set some generous upper
limits for two of the stars studied. Considering all our
abundance results, it is difficult to find an evolutionary status for
J-stars. Their average luminosity and variability types leads us to consider
these objects as less evolved than normal (N) carbon stars. However,
standard AGB models are unable to explain all their properties. On the contrary, 
the chemical peculiarities of J-stars suggests the existence of a non-standard
mixing mechanism, similar to that proposed in the red giant branch to 
explain anomalous CNO isotopic ratios and Li abundances. This extra-mixing
mechanism, acting preferably in the early AGB phase of low-mass
stars (M$\lesssim 2-3$ M$_\odot$), would take material from the
convective envelope, transport it down to regions hot enough for some
nuclear processing and then transport it back up to the convective
envelope. The expected stellar mass for the occurrence of this
cool processing  would be in agreement with the observational
evidence suggesting a low-mass for most of the J-stars studied
here. Mixing at the He-core flash and the binary system hypothesis 
may well be alternative scenarios, although fine tuning is required to explain 
all the observed characteristics of J-stars within these models. Nevertheless,
these scenarios require further investigation. 

On the other hand, the existence of rather luminous J-stars 
(M$_{\rm{bol}}<-5.5$) in our galaxy, as well as in other galaxies
(M31 and the Magellanic Clouds, see Brewer et al. 1996; Smith et al.
1995; Bessell, Wood and Lloyd-Evans 1983), suggests there could
be two types of J-stars depending upon the initial mass of the
parent star. The low-mass J-stars would be explained by a non-standard
mixing mechanism such as those mentioned above, while the high initial
mass (M$\gtrsim4$ M$_\odot$) J-stars (perhaps WZ Cas is an example of
this) would be explained through the operation of HBB. This idea
was previously proposed by \citet{lor96}. Note that current models
of s-process nucleosynthesis (e.g. Busso \& Gallino 1997) predict a strong
metallicity dependence of s-nuclei enrichment. For solar 
metallicity and/or slightly metal-rich intermediate mass AGB stars, a
small s-process element enrichment is predicted. This might be in
agreement with the small [$<h>$/M] value that we have found in WZ Cas.
The study of the presence of s-elements in the luminous J-stars of
M31 and the Magellanic Clouds would be of a great interest.

\acknowledgements

Data from the VALD data base at Vienna were used for the preparation
of this paper. K. Eriksson and the stellar atmosphere
group of the Uppsala Observatory are thanked for providing the grid
atmospheres. The 4.2 m WHT is operated on the island of La Palma by the
RGO in the Spanish Observatorio del Roque de los Muchachos of the
Instituto de Astrof\'\i sica de Canarias. Based in part on observations
collected at the German-Spanish Astronomical Centre, Carlar Alto, Spain.
This work was partially supported by grant PB96-1428.

\clearpage

\figcaption[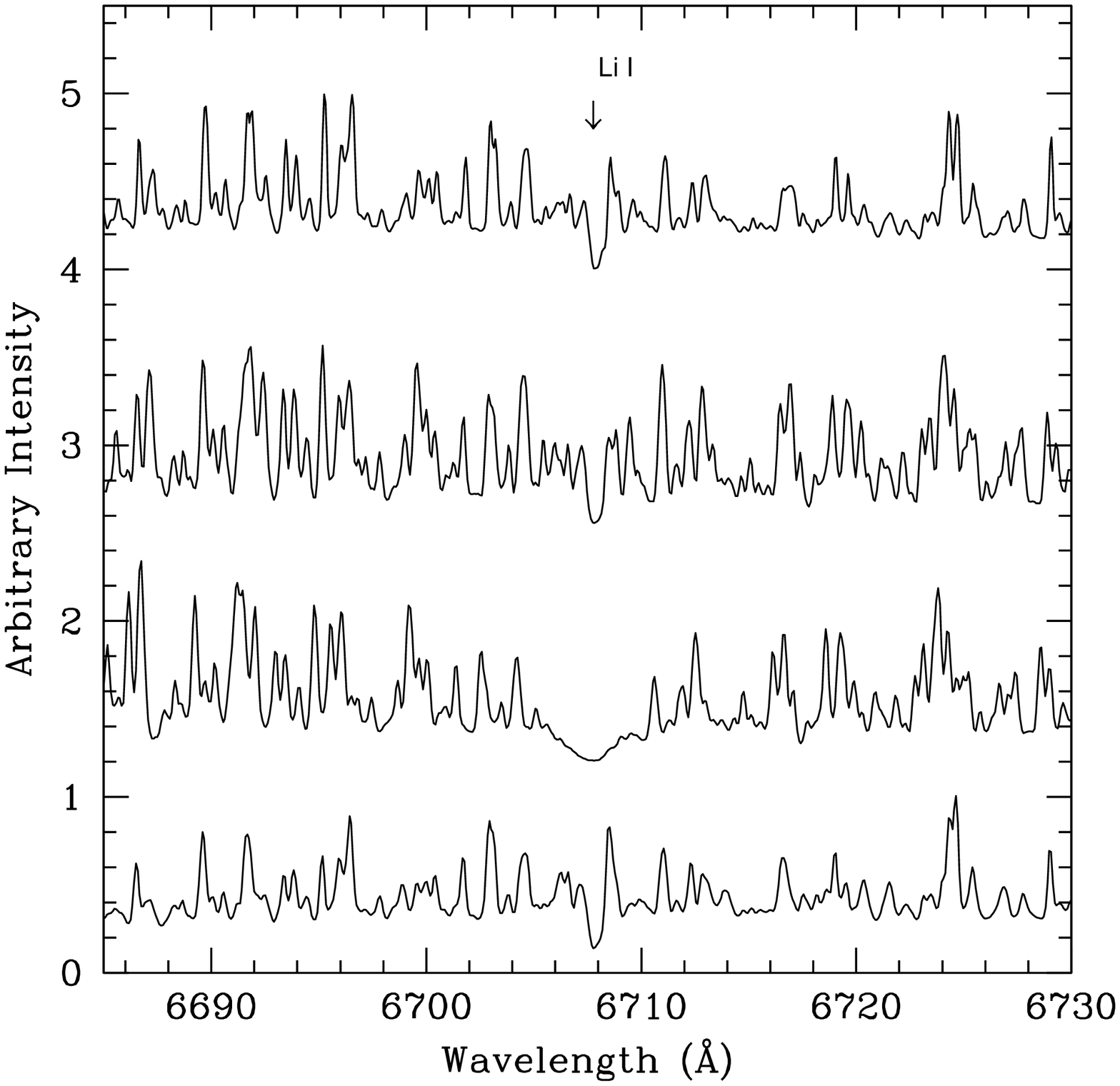]{Spectral region of the $\lambda 6708$ {\AA} Li I line of four 
J-stars in the sample; from top to bottom: RY Dra, UV Cam, WX Cyg and VX And. Note the
very strong Li absorption in WX Cyg. \label{fig1}} 

\figcaption[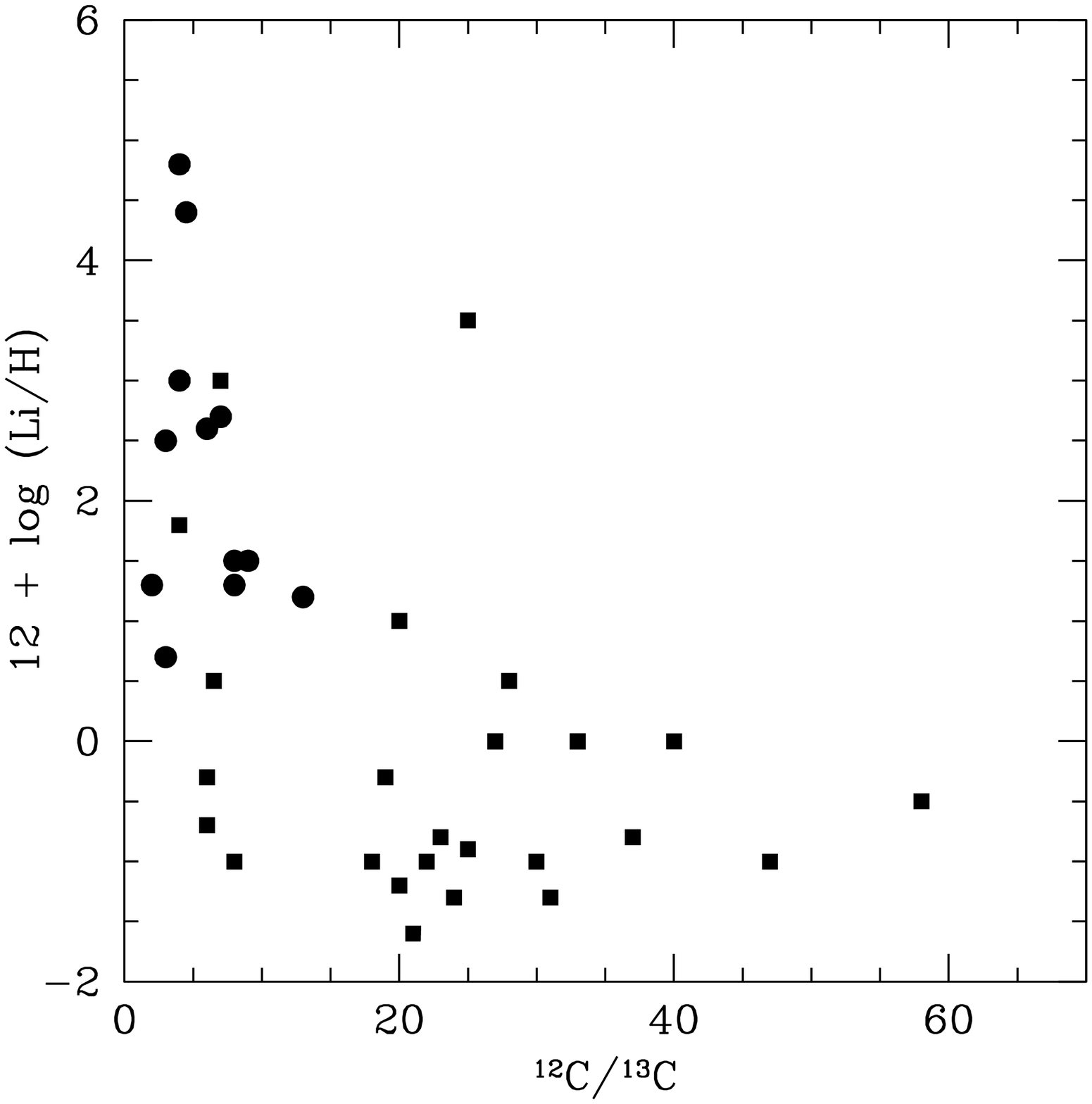]{Li abundances versus $^{12}$C/$^{13}$C ratios
in J-stars on this work (circles) and N-stars (squares) from
Abia \& Isern (1997). All the J-stars are Li-rich. Note that there
are some Li-rich N-stars with low carbon isotope ratios. \label{fig2}}

\figcaption[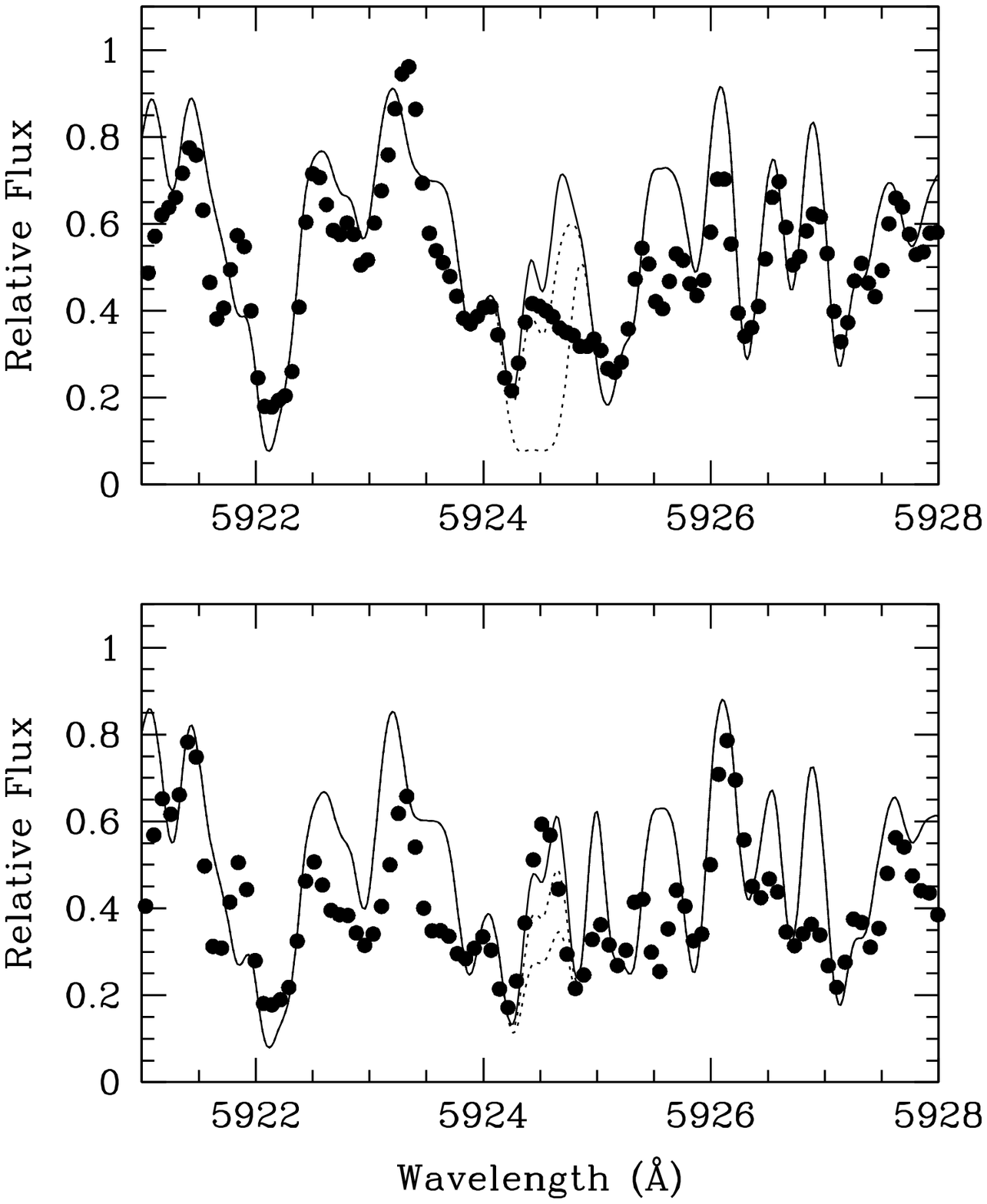]{Top panel:  Observed (dots) spectrum  of WZ Cas
and synthetic fits (continuous and dashed  lines) in the region of the
$\lambda 5924$  {\AA} Tc I line. The  fits shown are   for no Tc, log
$\epsilon$(Tc)=1  and 2.   Bottom panel:  The  same for   the  star UV
Cam.   Synthetic fits   are  for   no  Tc,  log $\epsilon$(Tc)=1   and
1.5. \label{fig3}}

\figcaption[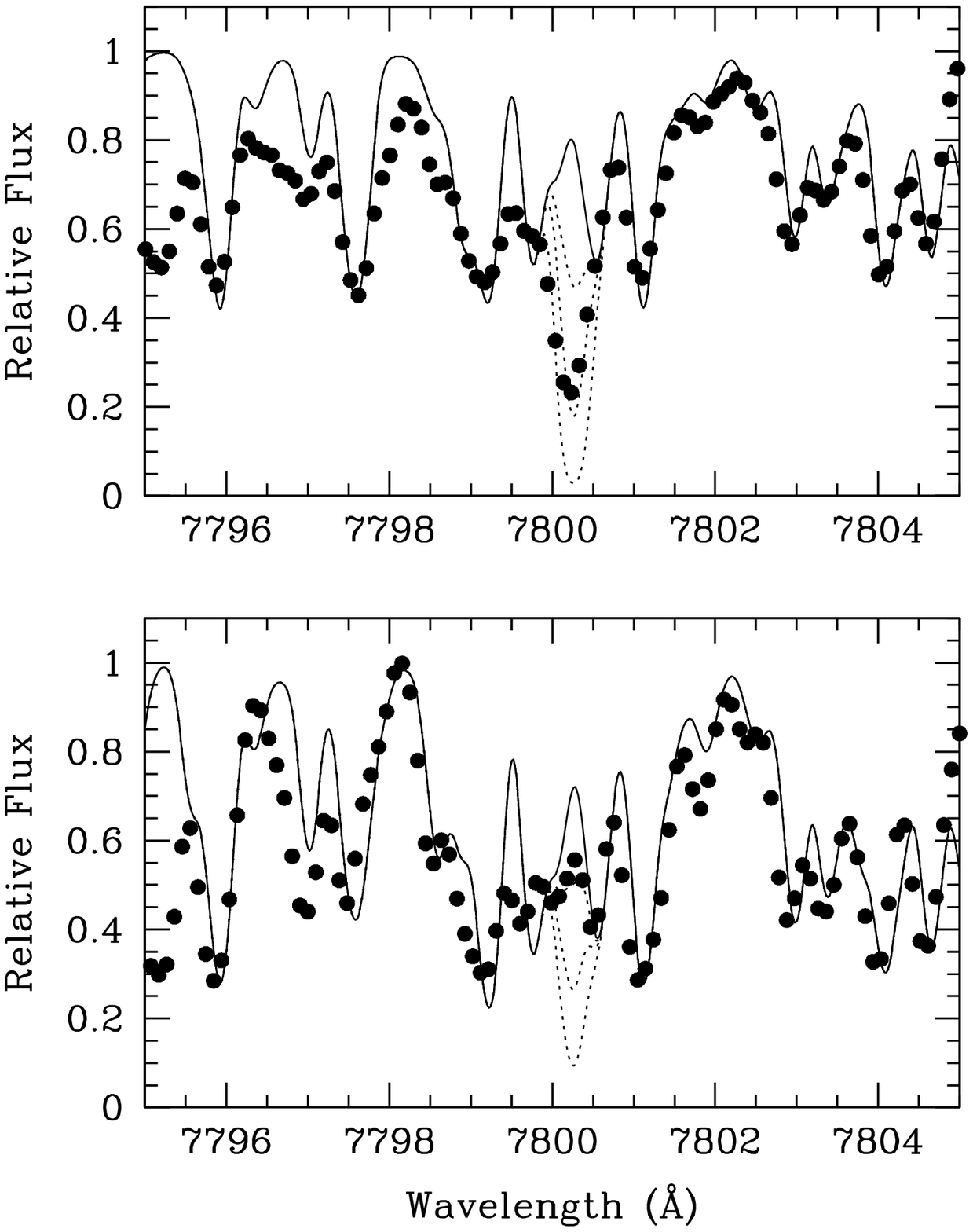]{Top panel:  Observed (dots) spectrum  of WZ
Cas and synthetic fits (continuous and dashed  lines) in the region of
the $\lambda 7800$ {\AA} Rb I line. The fits shown are for no Rb, log
$\epsilon$(Rb)=1.5, 2.1 and 2.6. Bottom panel: The same  for the star UV
Cam. Fits are for no Rb, log $\epsilon$(Rb)=1.5, 1.8 and 2.1. 
\label{fig4}}

\figcaption[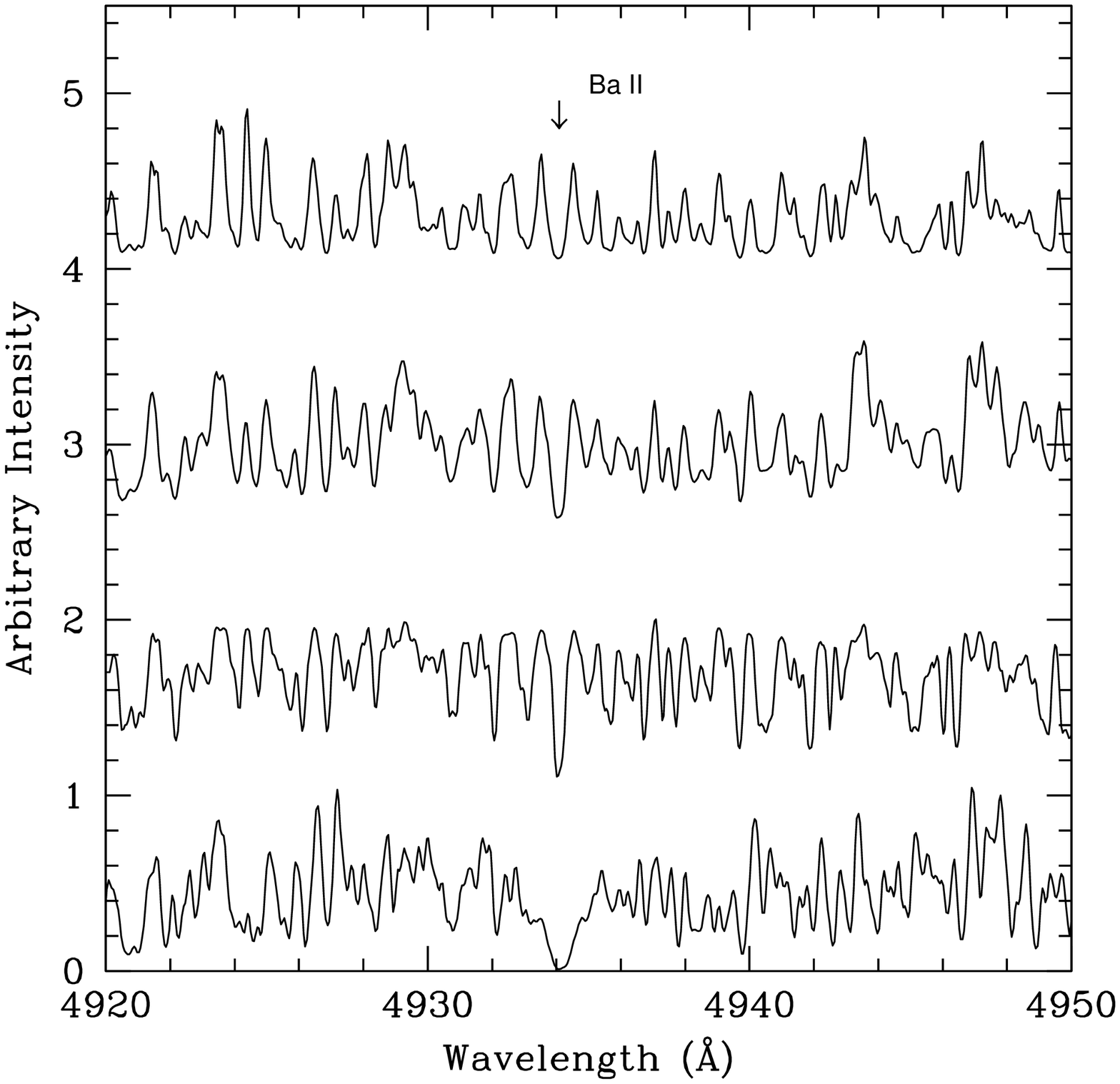]{Comparison of $\lambda 4934.07$ {\AA} Ba II line strengths  
in three J-stars (from top to  bottom: V614 Mon,  BM Gem and  Y CVn) and a normal
N-type carbon star Z Psc (bottom). \label{fig5}} 

\figcaption[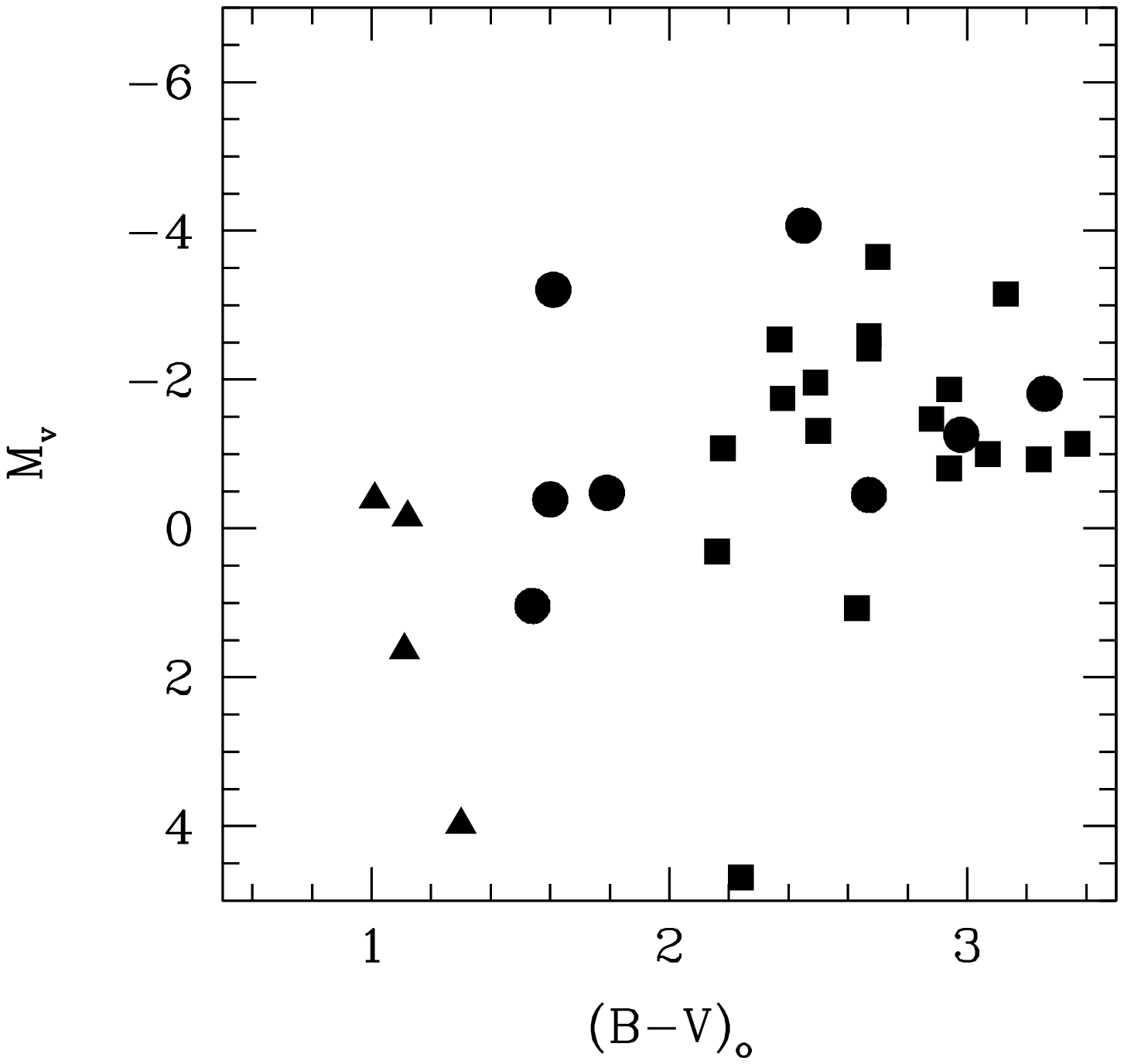]{Observational   H-R   diagram      for   carbon
stars.   (circles) J-type in    this study, (squares)   N-type stars  and
(triangles)  R-type stars.  Data for  N  and  R  stars are taken  from
Alksnis et al. (1998). \label{fig6}} 

\clearpage 
\plotone{fig1j.eps}
\clearpage
\plotone{fig2j.eps}
\clearpage
\plotone{fig3j.eps}
\clearpage
\plotone{fig4j.eps}
\clearpage
\plotone{fig5j.eps}
\clearpage
\plotone{fig6j.eps}

\end{document}